\documentclass[%
reprint,
superscriptaddress,
amsmath,amssymb,
aps,
prl
]{revtex4-1}

\usepackage{graphicx}
\usepackage{dcolumn}
\usepackage{bm}

\begin{document}

\preprint{APS/123-QED}
\title{Robust skyrmion-bubble textures in SrRuO$_3$ thin films stabilized by magnetic anisotropy}

\author{P. Zhang}
\email{p.zhang@rug.nl}
\author{A. Das}
\author{E. Barts}
\author{M. Azhar}
\affiliation{University of Groningen, Zernike Institute for Advanced Materials, 9747 AG Groningen, The Netherlands}
\author{L. Si}
\author{K. Held}
\affiliation{Institut für Festkörperphysik, TU Wien, Wiedner Hauptstraße 8-10, 1040 Vienna, Austria}
\author{M. Mostovoy}
\author{T. Banerjee}
\email{t.banerjee@rug.nl}
\affiliation{University of Groningen, Zernike Institute for Advanced Materials, 9747 AG Groningen, The Netherlands}

\date{\today}

\begin{abstract}

Topological spin textures in an itinerant ferromagnet, SrRuO$_3$ is studied combining Hall transport measurements and numerical simulations. We observe characteristic signatures of the Topological Hall Effect associated with skyrmions. A relatively large thickness of our films and absence of heavy-metal layers make the interfacial Dzyaloshinskii-Moriya interaction an unlikely source of these topological spin textures. Additionally, the transport anomalies exhibit an unprecedented  robustness to magnetic field tilting and temperature. Our numerical simulations suggest that this unconventional behavior results from magnetic bubbles with skyrmion topology stabilized by magnetodipolar interactions in an unexpected region of parameter space. 

\end{abstract}

\maketitle

The interplay between magnetism and electronic transport, enabled by spin-orbit coupling, has been explored to the study of different phenomena such as the anomalous Hall effect (AHE) \cite{karplus1954hall, smit1955spontaneous, berger1970side}, anisotropic magnetoresistance \cite{campbell1970spontaneous, smit1951magnetoresistance} and extrinsic and intrinsic spin Hall effect \cite{hirsch1999spin, d1971possibility, sinova2004universal}. Of them, the AHE has been widely pursued for over a century due to its rich physics in diverse magnetic systems. It originates from intrinsic and extrinsic contributions to quantum transport in magnetic materials 
i.e.\ the Berry curvature of the electronic bands and impurity scattering. 
The contribution of the AHE to electronic transport, namely in Hall resistivity, is represented as $\rho_{AHE}$=R$_s$M, relating it to the magnetization (M), while R$_S$ incorporates mechanisms that combine exchange interactions with spin-orbit coupling effects. More recently, topological Hall effect (THE), an additional component to the Hall resistivity \cite{ye1999berry, berry1984quantal, bogdanov2001chiral, soumyanarayanan2016emergent, bruno2004topological, bruno1999geometrically} resulting from an effective Lorentz force by a topological spin texture, has been  reported for systems with negligible and sizable spin-orbit coupling. The scalar chirality of non-coplanar spin structures, such as the magnetic skyrmion, gives rise to an effective magnetic field with a quantized flux proportional to the topological skyrmion number \cite{nagaosa2013topological, taguchi2003magnetic}.
Widely investigated in the context of the THE are the bulk chiral magnets as well as multilayers of conventional magnets with heavy-metal materials \cite{neubauer2009topological,muhlbauer2009skyrmion, pappas2009chiral, lee2009unusual, soumyanarayanan2017tunable}, offering novel physics with prospects of applications in non volatile memory devices. \\ 
The material class of correlated oxides has garnered recent attention in the exploration of skyrmions \cite{rowland2016skyrmions,banerjee2014enhanced} and exploiting the THE \cite{vistoli2019giant}, particularly in the itinerant ferromagnet SrRuO$_3$ (SRO) \cite{matsuno2016interface, meng2019observation, groenendijk2018berry}. Spin-orbit coupling has been exploited earlier, to the study of different emergent phenomena in SRO \cite{kats2004testing, roy2015intermixing, gupta2014strain, werner2008spin, roy2015interface, ziese2010structural}  and more recently to the study of the THE \cite{matsuno2016interface, meng2019observation, groenendijk2018berry} in devices utilizing SRO interfaced or capped with SrIrO$_3$ (SIO) and ascribed to skyrmion textures.
However, the understanding and origin of such magnetic textures that lead to the observation of THE is surprisingly lacking. For an itinerant ferromagnet such as SRO, it is expected, that the competition between different magnetic energies  such as magnetocrystalline anisotropy, long range magneto-dipolar forces and interfacial Dzyaloshinskii-Moriya interaction (DMI), when appropriately tailored can give rise to new magnetic textures. Despite this, such an interplay between magnetic energies in electronic transport was not addressed either in experimental or in theoretical studies.
This is quintessential not only for a comprehensive understanding of the origin of the THE but also for their stabilization and manipulation by applied currents or electric fields.\\
In this Letter, we report unconventional features in the Hall transport in SRO  films, akin to recent studies on THE in SIO/SRO bilayers \cite{matsuno2016interface, ohuchi2018electric} that were ascribed to skyrmions. Our films are tailored to be ferromagnetic and multidomain, thus expected to display a complex magnetocrystalline anisotropy dependence both with the applied magnetic field and temperature below the magnetic phase transition temperature of 120 K. We invoke the role of magnetostatic interactions, anisotropies and interfacial DMI to model the magnetic textures and explain our experimental findings by the formation of magnetic bubble domains with skyrmion topology. The robustness of the bubble arrays against the rotation of the magnetic field vector, unique to our findings, as well as the relatively high critical magnetic fields, are studied, and found to originate from the complex angular dependence of the magnetic anisotropy energy in our SRO films. \\
SRO, an archetypal member of the ruthenate family is an itinerant ferromagnet with large spin-orbit coupling and a complex magnetocrystalline anisotropy dependence \cite{kanbayasi1976magnetocrystalline, ziese2010structural, klein1996transport} with temperature below the T$_C$ of 150-160 K in bulk. Epitaxial SRO films were deposited on terminated and annealed (100) SrTiO$_3$ (STO) single-crystal substrates by pulsed laser deposition (PLD).
Reflection high-energy electron diffraction (RHEED) was used to monitor the growth of SRO thin films \textit{in-situ}.
All the films discussed in this work (A and B in the main text and C in the supplementary) \cite{Supplementary2020} are compressively strained with a lattice mismatch of 0.64${\%}$ with STO and 8.7 $\pm{0.2}$ nm thick as obtained from x-ray reflectivity (XRR) studies. Further information on the films are summarized in the supplementary information \cite{Supplementary2020} (Fig. S1, S2 and Table 1).
The atomic force microscopy (AFM) topology (Fig. S1) before and after the thin film deposition reveals the presence of both TiO$_2$ and SrO surface termination for all substrates used. Such double terminated substrates were found to result in local differences in structural and electronic properties at the different terminating sites, in our earlier works \cite{roy2015interface} and are an important consideration for this study.\\
Resistivity studies on unpatterned films (A and B) were done in standard four-terminal van der Pauw geometry for temperatures between 5 K to 300 K and shown in Fig. 1a. The differences in the temperature dependence of $\rho_{xx}$ for such thick SRO films (A and B), in spite of similar deposition conditions and thickness, are remarkable, and underscore the role of the local differences in substrate termination to electronic transport and ferromagnetic transition temperature (T$_c$=115 K for film A and 120 K for film B). The magnetization of the thin films was studied at variable temperatures by sweeping the applied magnetic field along the in-plane and out-of-plane directions using a Quantum Design superconducting quantum interference device (SQUID). One such measurement, for the as deposited film B, shown in Fig. 1b, is suggestive of multiaxial magnetocrystalline anisotropy  \cite{ziese2010structural, roy2015intermixing}. The easy axis of such strained thin films has a strong temperature-dependent angular variation with the film normal, leading to a complex magnetocrystalline anisotropy, as reported in earlier studies \cite {ziese2010structural, lindfors2017topological}.\\
Subsequently the films were patterned into a Hall bar geometry (Fig. 3a) using standard lithography techniques and ion beam etching for Hall transport studies. The transverse resistivity $\rho_{xy}$, can be written as the sum of the following contributions
$\rho_{xy}$ =$R_o B_\perp + R_s M_\perp (B_\perp) + \rho_{THE}$ and is antisymmetric in both $\mathbf{B}$ and $\mathbf{M}$. The bottom panel of Figs. 1 (c,d) shows the scaling of the anomalous hall resistivity ($\rho_{AHE}$) and conductivity ($\sigma_{AHE}$) for films A and B with temperature and magnetization respectively.  $\rho_{AHE}$ for all the films in this work have been obtained with an out-of plane magnetic field and at different temperatures as shown in Figs. 2b and 3b. 
$\rho_{AHE}$ is obtained from the zero-field resistivity after the subtraction of the ordinary hall background (details in \cite{Supplementary2020} and in Fig. S3).
A nonmonotonous dependence of $\rho_{AHE}$ and $\sigma_{AHE}$ is found for both samples with temperature upto T$_c$. Film A exhibits a sign reversal with temperature and magnetization, whereas Film B shows no such reversal. Such trends, typical of SRO films, indicate the intrinsic origin of the AHE associated with the intricacies of the temperature modulations in the band structure and their crossings at the Fermi energy \cite{ye1999berry,bruno2004topological, kats2004testing, roy2015intermixing}.  The differences in the trend, on the other hand, between the two films, are a manifestation of the local changes in the crystalline lattice structure and their corresponding influence on the band structure.\\ 
\begin{figure}[t]
 \centering
 \includegraphics[width=0.9\linewidth]{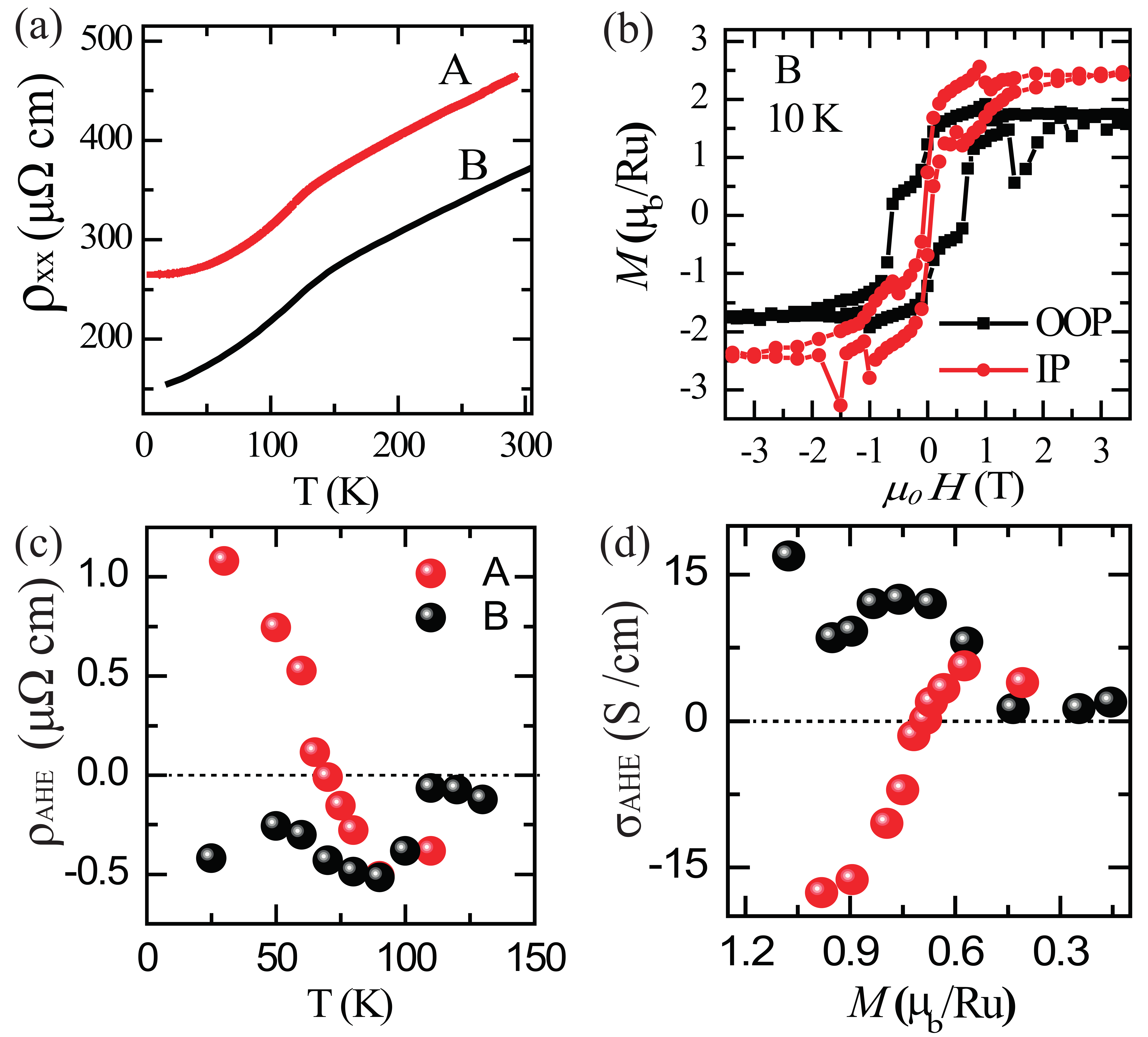}
 \caption{(a) Temperature dependence of the resistivity of SRO thin films deposited on nominally different substrate terminations but of similar thickness. (b) M-H curves measured at 10 K for Film B in out-of-plane (black) and in-plane (red) directions respectively. (c) Temperature dependence of anomalous Hall resistivity ($\rho_{AHE}$) for films A and B. Film A shows a sign change in $\rho_{AHE}$ close to T$_c$, Film B does not exhibit such a change and $\rho_{AHE}$ vanishes close to T$_c$. (d) Scaling of Anomalous Hall conductivity with magnetization for Film A (red) and B (black). A similar trend as in 1c with decreasing magnetization (on approaching T$_c$) is seen for film A.}
 \label{AHE}
 \end{figure}
A comprehensive study of the temperature dependence of the magnetization were carried out for both in plane and out of plane configurations and shown in Figure 2a is for Film A. The temperature dependence of the magnetization shows the evolution of a complex magnetocrystalline anisotropy in SRO with temperature, as discussed in literature \cite{ziese2010structural,lindfors2017topological} and consistent with the occurrence of an easy plane anisotropy.
Figure 2b and Fig 3b shows the magnetic field dependence of the electronic transport in films A and B respectively, patterned into a Hall bar with channel width of 50 $\mu$m. The direction of the applied field is shown in Fig. 3a. The data displayed in the panel is obtained after subtracting the component of the ordinary Hall effect. The subtracted negative ordinary Hall background confirms the carriers to be of n-type. Distinct features can be seen – the atypical AHE contribution corresponding to the magnetization of the films and its temperature dependence and the enhancement in transverse resistivity, $\rho_{xy}$, at higher fields manifested as humps in the $\rho_{xy}$ loops. The AHE signal is generally assumed to be proportional to the magnetization, obtained from the M-H loops, enabling the extraction of the THE contribution. However, a direct correlation  of the coercive fields H$_{c}$ obtained from bulk magnetization studies and from AHE are non-trivial and most often, not identical.\\  
\begin{figure}[b]
\centering
	\includegraphics[scale=0.085]{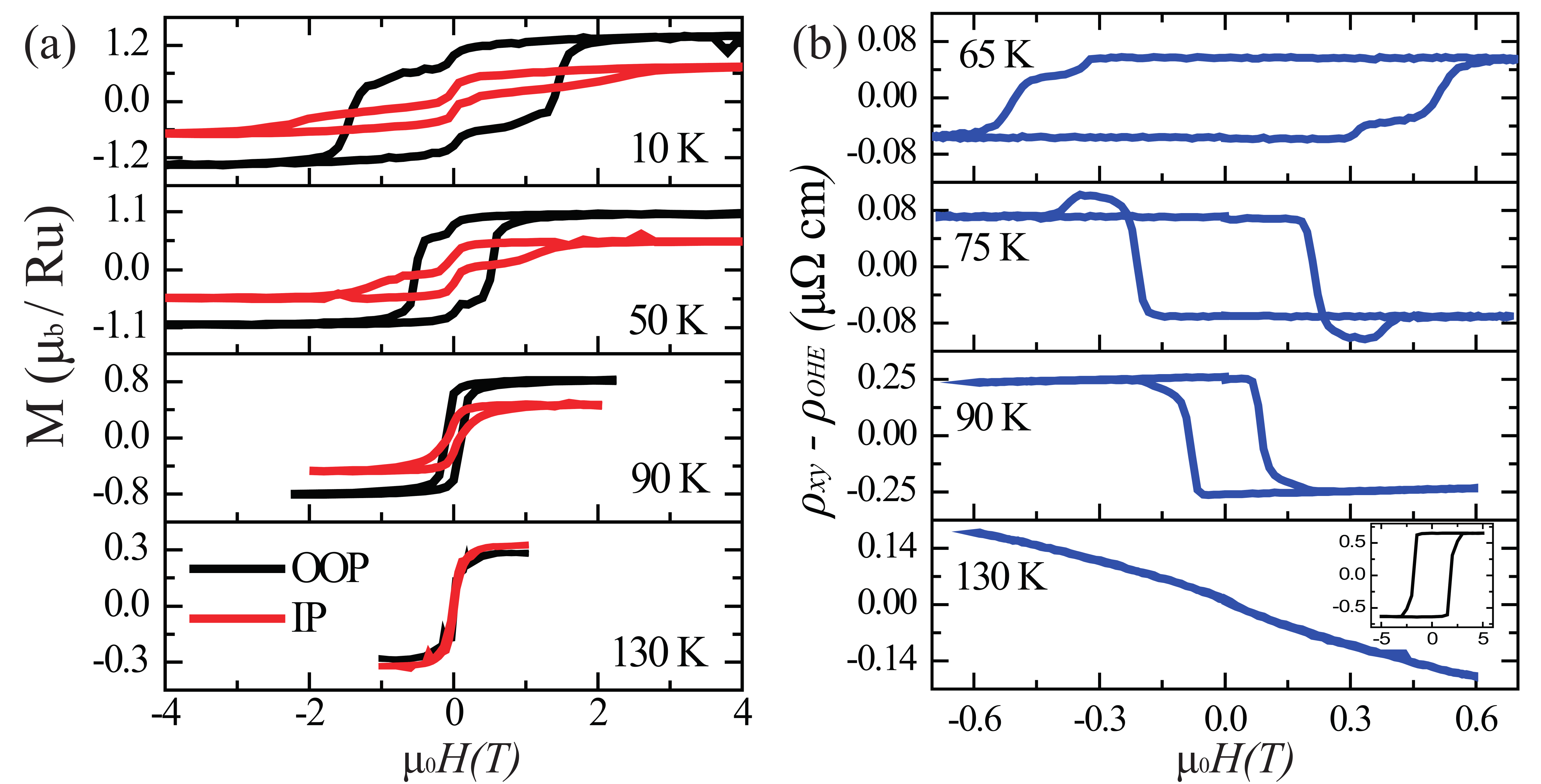}
 \caption{Temperature dependence of M-H curves and Hall resistivity for Film A. (a) M-H for out-of-plane (black) and in-plane (red) at different temperatures. Between each measurement, the film was always warmed to 300 K and thereafter cooled to the temperature of measurement. (b) Magnetic-field-dependent Hall resistivity measured at different temperatures. The linear contribution from OHE has been subtracted (except for 130 K). The inset shows the Hall resistivity at 10 K which changes sign beyond 70 K.}
 \label{Hall}
 \end{figure}
We further observe additional humps in $\rho_{xy}$ upto the magnetic phase transition temperature (T$_c$) for the films studied (Figs. 2b and 3b,c,d) and that are commonly ascribed to the THE, originating from the skew-scattering off an effective magnetic field induced by non-coplanar topological spin textures. The humps appear irrespective of the differences in either the sign of the AHE or in the magnetocrystalline anisotropy in Films A and B.
The THE in all cases is observed above 30 K and persists upto the T$_C$ of the films, while being remarkably robust to a large tilting angle with the applied field as shown in Figs 3c and d. These unique findings, reported for the first time in a non-chiral and relatively thick SRO films, imprint signatures of the complex temperature dependence of the multiple uniaxial anisotropies intrinsic to SRO and are associated with skyrmion bubbles. Upon increasing the magnetic field above 0.25 T, the THE effect was found to vanish in both films, corresponding to the collapse of topological spin textures as the magnetization reaches saturation, as shown in Fig. 2b and 3b.\\
\begin{figure}[t]
\centering
\includegraphics[scale=0.45]{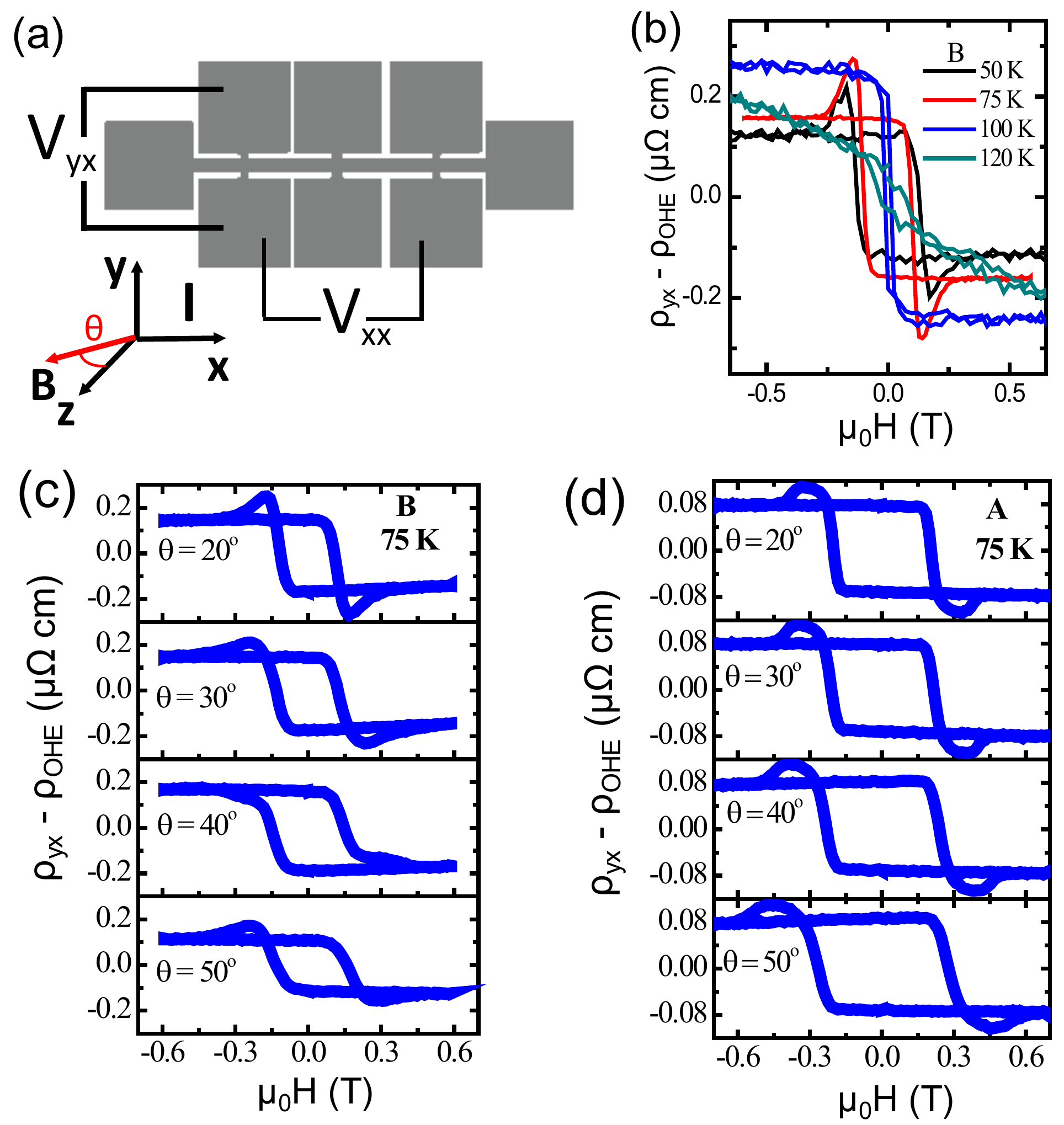}
\caption{Field and angle dependent Hall resistivity of SRO thin films. (a) Hall bar schematic showing longitudinal (V$_{xx}$) and transverse contact voltage contacts (V$_{yx}$). The tilt angle, $\theta$, with the magnetic field, B, in the out-of-plane direction and the current (I) along x-direction are also shown. (b) Field  dependent  Hall  resistivity at  different temperatures  for  Film  B. The linear contribution from OHE has been subtracted (except for 120 K). (c)(d)  Angular  dependent  Hall resistivity at 75 K for Film B and A respectively.}
\label{angle}
\end{figure}
To understand the occurrence of THE and the stability of the magnetic bubble array in our films, we performed numerical simulations, with the simplifying assumption that the magnetization vector in thin SRO film is independent of the vertical coordinate, $z$:   $\displaystyle{\bm{M}}=\bm{M}(x,y)$ \cite{garel1982phase}.
The energy of the film of thickness $h$ is,
\begin{equation}
\label{eq:mod}
\begin{split}
E &= 
h\int \!\! d^2 x \biggl[\,
\sum_{i=x,y}\frac{c}{2} (\partial_i \bm{M} )^2 
-\frac{K_1}{2} M_{z}^2 - \frac{K_2}{4 M_{\rm s}^2}M_{z}^4 
\\&
- \bm{H} \cdot \bm{M}+\frac{\lambda}{4 M_{\rm s}^2}\left( \bm{M}^2 -M_{\rm s}^2 \right)^2
\biggl]
+ E_{\rm ms}, 
\end{split}
\end{equation}
where the first term is the exchange energy, the second and third terms are the magnetocrystalline anisotropy energies of second and fourth order, respectively,  and the fourth term is the Zeeman energy (in CGS units). Rather than considering Landau expansion in powers of $M$ \cite{garel1982phase,yu2012magnetic}, we use the fifth term in Eq.(\ref{eq:mod}) with the parameter $\lambda = 100$ to constrain the magnitude of the magnetization to its saturation value, $\bm{M} = M_{\rm s} \bm m$ ($\bm m^2 = 1$). The energy of magnetostatic interactions, $E_{\rm ms}$, has a compact form in the reciprocal space~\cite{yu2012magnetic}. We minimize energy on the space of 61 Fourier harmonics of the magnetization (Fig.~\ref{fig:states} d) \cite{Supplementary2020}.
\begin{figure}[h!]
	\includegraphics[scale=0.7]{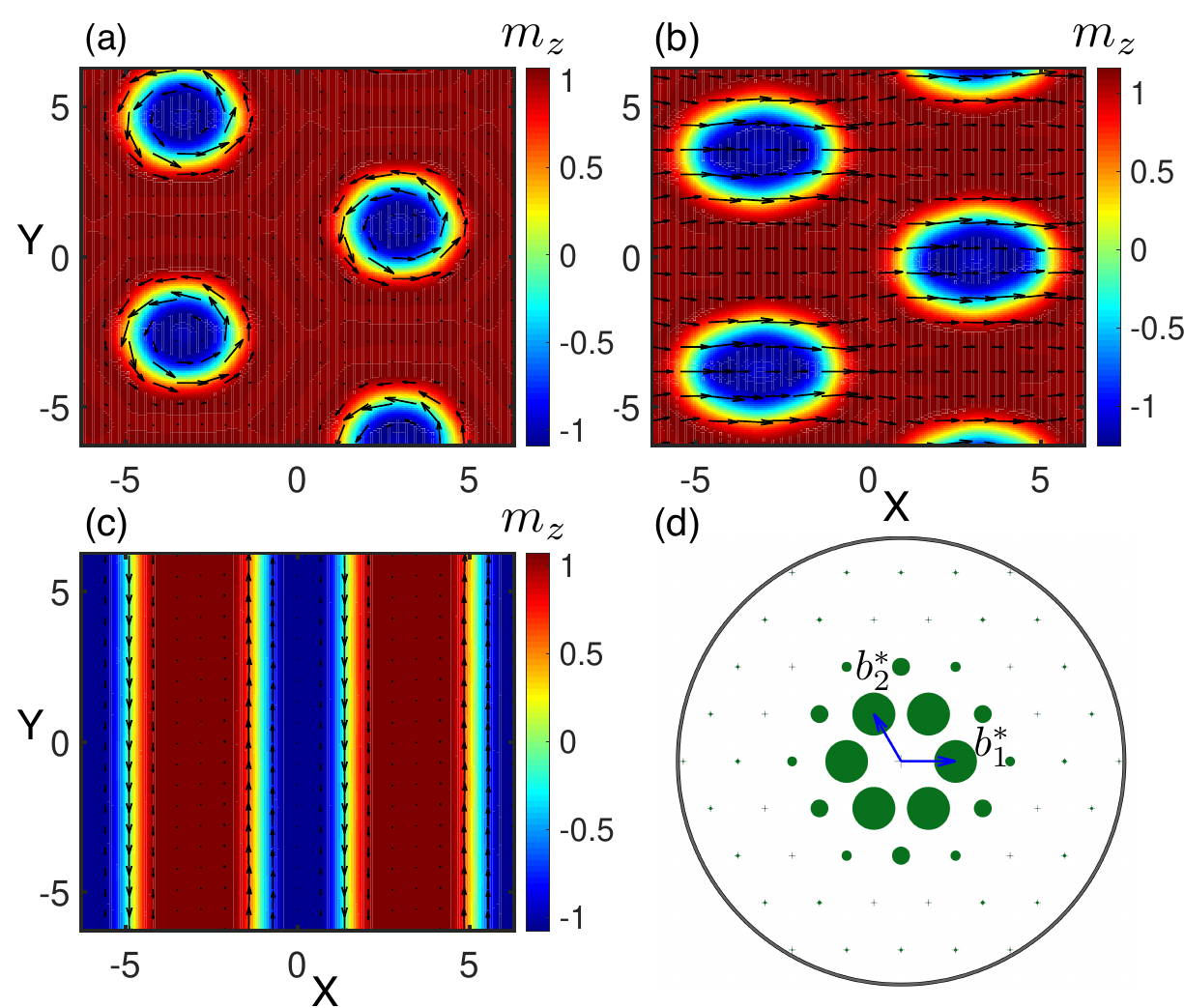}
	\caption{Competing non-collinear magnetic phases: (a) an array of bubbles with topological charge $-1$, (b) an array of bubbles carrying zero topological charge and an in-plane magnetic moment, and (c) the stripe domain state. In-plane components of $\bm m$ are shown with arrows, $m_z$ is color-coded.  Distances are given in units of the film thickness, $h$. (d) Skyrmion crystal in reciprocal space (large zero harmonic is excluded for clarity). The green dot area is proportional to the magnitude of the Fourier harmonic of the magnetization. Black line encircles the subspace of 61 wave vectors.
	}
	\label{fig:states}
\end{figure}

\begin{figure}[h!]
		\includegraphics[scale=0.7]{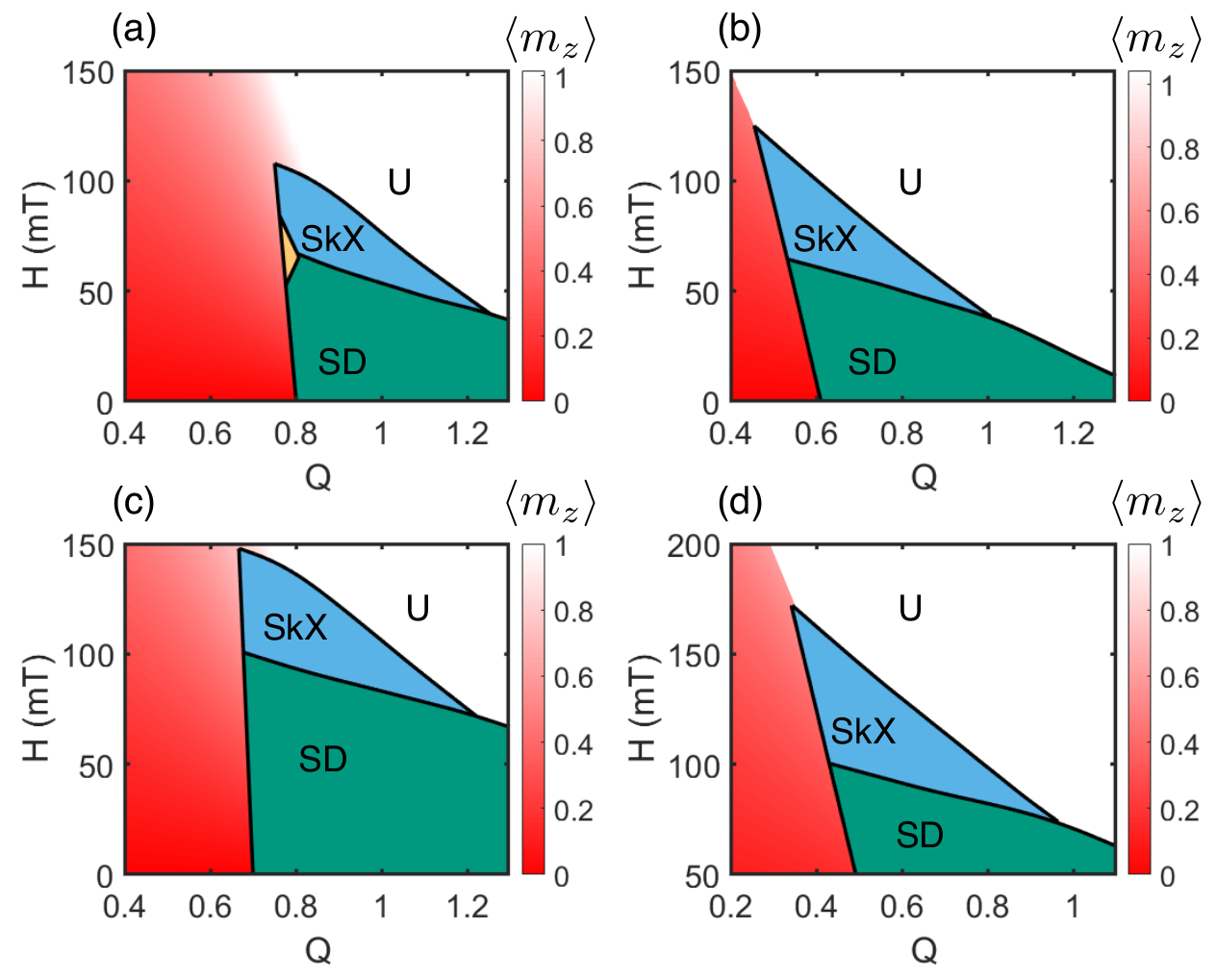}
	\caption{Magnetic field, $H$, vs quality factor, $Q = \frac{K_1}{4\pi}$, phase diagrams, which include the skyrmion crystal (SkX), stripe domain (SD) and uniform (U) states. Red color intensity indicates $m_z$ in the uniform state with a tilted magnetization. The state with the magnetization normal to the film is shown with white color. These diagrams are calculated for (a) $R =
 \frac{K_2}{4\pi} = 0$ and (b) $R =  0.4$. The interfacial DM interaction with $D=1$ widens the regions occupied by the SkX and SD phases,  calculated for (c) $R = 0$ and (d) $R =  0.4$.}
	\label{fig:pdQH}
\end{figure}

\begin{figure}[h!]
		\includegraphics[scale=0.7]{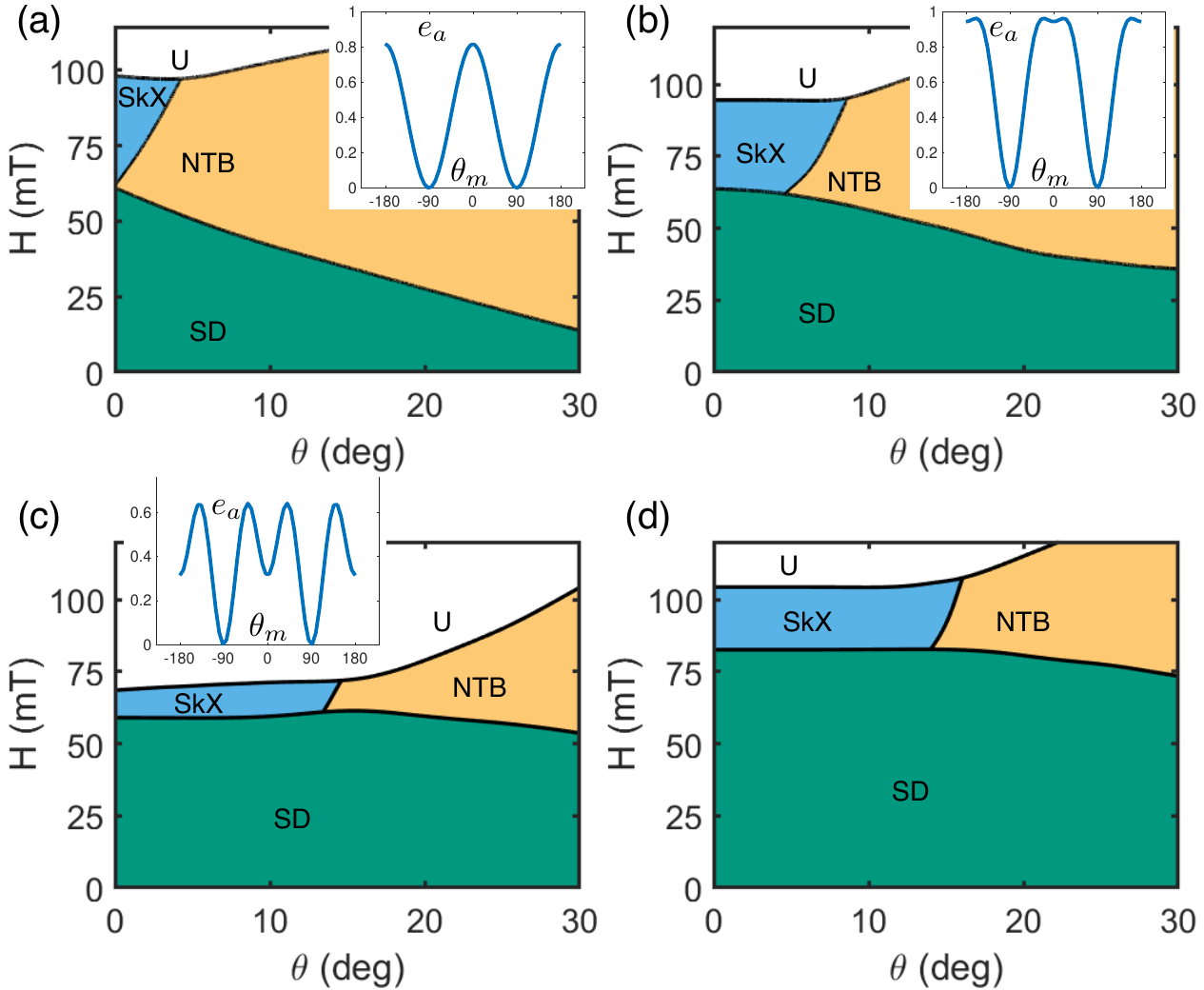}
	\caption{Phase diagrams in a tilted magnetic field, $\theta$ being the tilt angle, for (a) $Q=0.87$ and $R=0$, (b) $Q=0.65$ and $R=0.40$ and (c) $Q=0.65$ and $R = 0.60$, and (d) $Q=0.65, R = 0.60$ and $D=1$. In addition to the skyrmion crystal (SkX, blue), stripe domain (SD, green) and uniform (U, white) states,   these diagrams include an array of non-topological bubbles (NTB, yellow), shown in Fig.~\ref{fig:states}b. The insets show the dependence of the dimensionless anisotropy energy density, $e_{\rm a}$, in the uniform state on the magnetization tilt angle, $\theta_{\rm m}$,  for the corresponding parameter sets.}
	\label{fig:pdHth}
\end{figure}

\begin{figure}[!h]
 \includegraphics[width=0.8\linewidth]{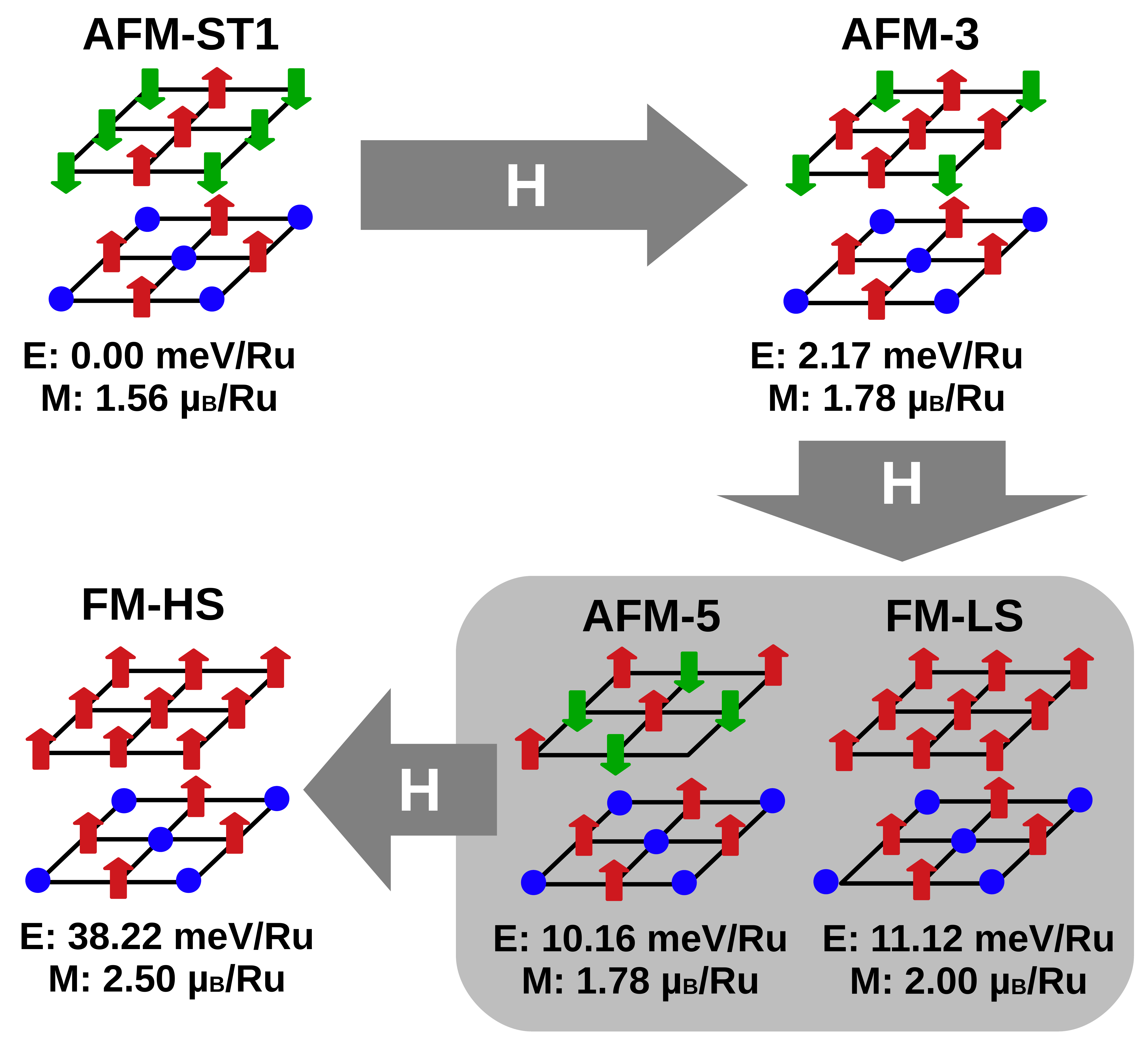}
\caption{Schematic picture of ferromagnetic (FM) and antiferromagnetic
(AFM) phases and  transitions in an external magnetic field as
calculated by DFT+$U$ (further details in \cite{Supplementary2020}). The $\uparrow$- and $\downarrow$-spins are on the Ru
sites, the (blue) circles are non-magnetic Ti sites (shown are only the Ru-Ti intermixed and the neighbouring Ru layer). E is the zero field energy of the phase and the M magnetic moment per Ru atom. The gray box indicates the possible coexistence of the AFM-5 and FM-LS
phase.\label{Fig:DFTU}}
\end{figure}

Figure~\ref{fig:states} shows noncollinear magnetic states stabilized by magnetostatic interactions and an applied magnetic field: a triangular array of Bloch skyrmions (type-I bubbles) with the topological charge $-1$ in real space (panel a) and in the reciprocal space (panel d), an array of type-II bubbles with zero topological charge favored by tilted magnetic fields (panel b) and the stripe domain state appearing in weak applied fields (panel c). 
Earlier studies of relatively thick films, suggested that stability of the stripe domain state and bubble array requires the quality factor, $Q = \frac{K_1}{4 \pi} > 1$~\cite{malozemoff1979magnetic,hubert1998magnetic}. For $Q  < 1$, the perpendicular magnetic anisotropy is not strong enough to overcome magnetostatic interactions favoring a uniform state with an in-plane magnetization. However, recent micromagnetic simulations of thin films showed that inhomogeneous magnetic states can be induced by an applied magnetic field even for $Q<1$ ~\cite{vousden2016skyrmions}. This is partly related to the fact that in thin films the domain wall width is no longer negligibly small compared to the film thickness, as can be seen from Fig.~\ref{fig:states}. In addition, the notion of an effective magnetic anisotropy, $K_{\rm eff} = K_1 - 4 \pi$, is only meaningful for uniform states, since dipole-dipole interactions strongly depend on the magnetic modulation wave vector. 
Moreover, the phase diagram in Fig.~\ref{fig:pdQH}a calculated for the magnetic field normal to the film, shows that the field interval in which the skyrmion crystal (SkX) has the lowest energy widens as the quality factor decreases. The SkX becomes unstable for $Q \lesssim 0.75$, near the line separating the uniform states with the perpendicular (white color) and tilted (red color) magnetization.  The stability region of the SkX phase extends to lower $Q$ and higher critical fields when we include the 4$^{\rm th}$-order  magnetocrystalline anisotropy, which is relatively large in SRO  due the strong spin-orbit coupling of Ru \cite{kanbayasi1976magnetocrystalline, klein1996transport}. Figure ~\ref{fig:pdQH}b shows the phase diagram calculated for $R = \frac{K_2}{4 \pi} = 0.4$.  $R > 0$ favors magnetization normal to the film, which makes the SkX more stable.
Thus a smaller quality factor in combination with a higher-order anisotropy can significantly increase stability of the SkX, although the fields at which the SkX undergoes transition into a uniform state (Fig.~\ref{fig:pdQH}) are still 2-3 times lower than the field of $\sim 400$ mT  at which the THE disappears in experiment.
In tilted magnetic fields (see Fig. 3a) the bubbles with skyrmion topology become unstable and transform into non-topological  bubbles (NTB) with two Bloch points (Fig.~\ref{fig:states}b) and an in-plane magnetic dipole moment, resulting in disappearance of the THE. In the M-type hexaferrite with $Q \sim 1$, the topological phase  disappears at a small tilt angle, $\theta = 2.3^\circ$~\cite{yu2012magnetic}. The THE found at much larger $\theta$ in our experiment can be explained by strong 4$^{\rm th}$-order anisotropy
(see Fig.~\ref{fig:pdHth}). The insets show angular dependence of the anisotropy energy in the uniform state as a function of the magnetization tilt angle, $\theta_{\rm m}$. As $R$ increases, the anisotropy energy, in addition to the global minimum at $\theta_{\rm m} = \pm \frac{\pi}{2}$ (in-plane magnetization), acquires a local minimum at $\theta_{\rm m} = 0,\pi$ (out-of-plane magnetization), which leads to an increase of the field tilt angle $\theta$, at which the THE disappears.
The interfacial Dzyaloshinskii-Moriya interaction, 
\begin{equation}
\label{eq:DMI}
E_{\rm DM} = 
D h^2 \int \!\! d^2 x \, 
\biggl[ 
(\bm{M}\cdot \bm{\nabla}) M_z - M_z (\bm{\nabla} \cdot \bm{M})
\biggl].
\end{equation}
is effectively small due to the large number of magnetic RuO$_2$-layers in the SRO film and the absence of heavy-metal elements. However, it can make skyrmions more resilent to high magnetic fields (see Figs.~\ref{fig:pdQH}~c,d), as they acquire a N\'eel component (Fig. S6 in \cite{Supplementary2020}).  In addition, bubbles with skyrmion topology become more stable against the transition into non-topological bubbles in tilted magnetic fields, as shown in Figs.~\ref{fig:pdHth}~c,d. 
Finally, we address the differences in the magnetic properties of the films deposited under similar conditions. These are clearly observed in the remnant and saturation magnetizations in film A and B (Fig. 1b and Fig 2a). Films A and B further display two consecutive transitions below 70 K with magnetic field that is representative of a metamagnetic behavior. Using DFT+$U$, we can explain the differences between films A and B as well as the two consecutive transitions by considering the differences in the Ti intermixing at the SRO/STO interface.  
We find that Ti intermixing stabilizes different magnetic phases with similar energies, see Fig.~\ref{Fig:DFTU} and supplementary material \cite{Supplementary2020}, including a FM high spin (HS) state with a magnetic moment of 2.5~$\mu_{\rm B}$/Ru as experimentally observed for Film B (Fig. 1b). Note that this magnetic moment is larger than the bulk magnetic moment [2~$\mu_{\rm B}$/Ru corresponding to the FM low spin (LS) state in Fig.~\ref{Fig:DFTU}]. The first staggered AFM-ST1 state has a magnetic moment of 1.56~$\mu_{\rm B}$/Ru. The  DFT+$U$ calculated magnetic moments and transitions agree well with our key experimental observations. \\ 
The observation of THE in thick non-chiral SRO films, in the absence of heavy-metal layers, is ascribed to robust magnetic bubble domains with skyrmion topology and explained by incorporating the hitherto ignored contributions of the second- and fourth-order magnetocrystalline anisotropy terms to the total energy and for $Q<1$. We find that the robustness of such bubble domains against the rotation of the magnetic field vector is significantly influenced by the complex angular dependence of the multiaxial anisotropy energy in SRO films. These considerations were surprisingly lacking in recent studies \cite{groenendijk2018berry,meng2019observation,ohuchi2018electric,matsuno2016interface} carried out in relatively thinner SRO films using AHE and magnetic force microscope techniques. We believe that a direct quantitative evaluation of the magnetization vector and interpretation of the magnetic domains in thick SRO films, possessing strong temperature dependence of the multiaxial anisotropy, with reasonable spatial and temporal resolution, is challenging using magnetic force microscopy measurements. Such a non-trivial competition between different local and non-local magnetic energies by designing magnetocrystalline anisotropies at engineered interfaces, opens new opportunities for their manipulation by electric fields and spin orbit fields for diverse applications in oxide based spintronics.

\begin{acknowledgments}
PZ thanks China Scholarship Council, MA and AD acknowledges Dieptestrategie grants (2014 and 2016),  Zernike Institute for Advanced Materials, for financial support. PZ, AD and TB acknowledges J. Baas, J. G. Holstein and H. H. de Vries for technical assistance and A. A. Burema, S. Chen and J. J. L. van Rijn for discussions. TB acknowledges discussions with M. A. Frantiu on micromagnetic simulations. This work was realized using NanoLab-NL facilities.  LS and KH have been supported by the Austrian Science Fund (FWF) through project P30997. EB and MM acknowledge Vrije FOM-programma `Skyrmionics'.
\end{acknowledgments}


%

\end{document}